\documentclass[reprint,showpacs,preprintnumbers,amsmath,a4paper,amssymb,xcolor=dvipsnames]{revtex4-1}
\pdfoutput=1

\usepackage[pdftex]{graphicx}
\usepackage{graphicx}
\usepackage{dcolumn}
\usepackage{bm}
\usepackage{datetime}
\usepackage{float}
\usepackage[pdftex,colorlinks=true,urlcolor=black,linkcolor=black]{hyperref} 
\usepackage{microtype}
\usepackage{epstopdf} 
\usepackage{siunitx}
\usepackage{amsmath}
\usepackage{amssymb}
\usepackage{braket}
\usepackage{xcolor}

\newcommand{\Note}[1]{\textcolor{black}{#1}}

\newcommand{\NoteV}[1]{\textcolor{black}{#1}}
\newcommand{\NoteVV}[1]{\textcolor{black}{#1}}
\DeclareSIUnit\micron{\micro\metre}
\DeclareSIUnit\mrad{\milli\rad}
\DeclareSIUnit\gauss{G}
\DeclareSIUnit\photons{photons}
\DeclareSIUnit\pixel{px}
\DeclareSIUnit\er{\textit{E}_R}

\begin{document}

\title{Spin-resolved single-atom imaging of $^6$Li in free space}

\author{Andrea Bergschneider}
	\email{bergschneider@physi.uni-heidelberg.de;}
	\thanks{ \\ Present address: Institut f\"ur Quantenoptik und Quanteninformation (IQOQI), \"Osterreichische Akademie der Wissenschaften, 6020 Innsbruck, Austria}
	\affiliation{Physikalisches Institut der Universit\"at Heidelberg, Im Neuenheimer Feld 226, 69120 Heidelberg, Germany}

\author{Vincent M. Klinkhamer}
	\email{klinkhamer@physi.uni-heidelberg.de}
	\affiliation{Physikalisches Institut der Universit\"at Heidelberg, Im Neuenheimer Feld 226, 69120 Heidelberg, Germany}
\author{Jan Hendrik Becher}
	\affiliation{Physikalisches Institut der Universit\"at Heidelberg, Im Neuenheimer Feld 226, 69120 Heidelberg, Germany}
\author{Ralf Klemt}
	\affiliation{Physikalisches Institut der Universit\"at Heidelberg, Im Neuenheimer Feld 226, 69120 Heidelberg, Germany}
\author{Gerhard Z\"urn}
	\affiliation{Physikalisches Institut der Universit\"at Heidelberg, Im Neuenheimer Feld 226, 69120 Heidelberg, Germany}
\author{Philipp M. Preiss}
	\affiliation{Physikalisches Institut der Universit\"at Heidelberg, Im Neuenheimer Feld 226, 69120 Heidelberg, Germany}
\author{Selim Jochim}
	\affiliation{Physikalisches Institut der Universit\"at Heidelberg, Im Neuenheimer Feld 226, 69120 Heidelberg, Germany}
	

\begin{abstract}
We present a versatile imaging scheme for fermionic  $^6$Li atoms with single-particle sensitivity. Our method works for freely propagating particles and completely eliminates the need for confining potentials during the imaging process. We illuminate individual atoms in free space with resonant light and collect their fluorescence on an electron-multiplying CCD camera using a high-numerical-aperture imaging system. We detect approximately \num{20} photons per atom during an exposure of \SI{20}{\us} and identify individual atoms with a fidelity of  \SI{99.4\pm0.3}{\percent} . By addressing different optical transitions during two exposures in rapid succession, we additionally resolve the hyperfine spin state of each particle. The position uncertainty of the imaging scheme is  \SI{4.0}{\micron}, given by the diffusive motion of the particles during the imaging pulse. The absence of confining potentials enables readout procedures, such as the measurement of single-particle momenta in time of flight, which we demonstrate here. Our imaging scheme is technically simple and easily adapted to other atomic species.
\end{abstract}

\maketitle

\section{Introduction}

Studying the microscopic mechanism of quantum phenomena requires probing of observables on a single-particle level. With ultracold atomic systems such single-particle resolved detection schemes are possible. This not only makes them excellent candidates for quantum simulation but also enables the extraction of higher-order correlation functions which hold a central place in the description of many-body quantum phenomena.

In recent years, single-atom imaging techniques have been developed for numerous ultracold atom systems \cite{Ott2016}. A notable example is the so-called quantum gas microscope scheme \cite{Nelson2007, Karski2009, Bakr2009, Sherson2010, Cheuk2015,  Haller2015, Parsons2015, Omran2015,  Edge2015, Miranda2015, Yamamoto2016} where atoms that initially occupy the potential wells of an optical lattice 
are pinned to the corresponding sites and imaged based on the collection of fluorescence photons with a high-resolution objective. To prevent the atoms from hopping to adjacent sites during the imaging process, elaborate cooling schemes have to be applied to colle	ct hundreds of photons per atom. However, these schemes typically lead to the loss of the spin information in the system. Furthermore, pinning and imaging the atoms in a deep trap causes light-assisted collisions which project the initial occupation number per site to odd and even values. Therefore, complex experimental schemes have been developed to retrieve spin and occupation number information \cite{Gibbons2011, Fuhrmanek2011b, Preiss2015, Boll2016}.

\begin{figure}
        \centering
               \includegraphics[width=\columnwidth]{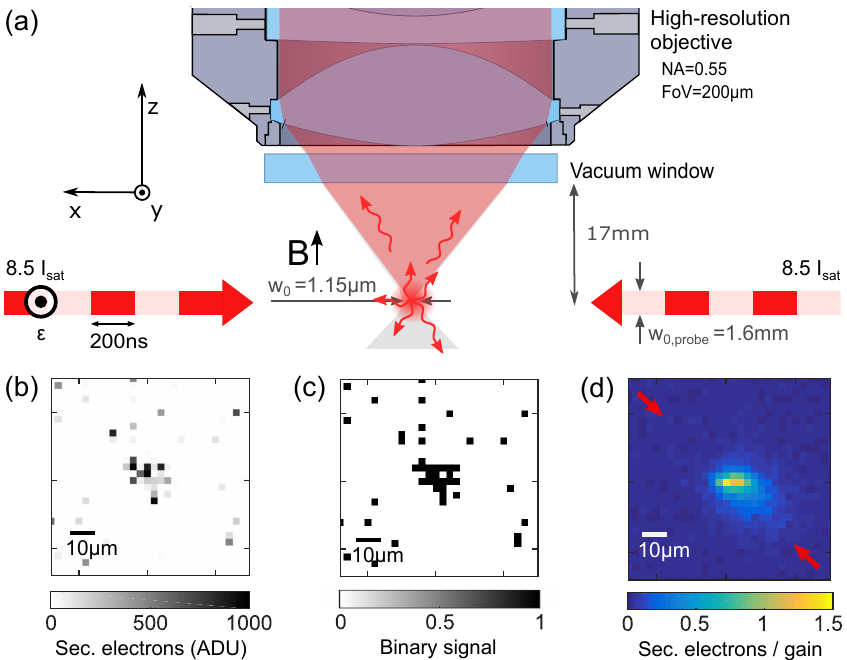}
        \caption{{Single-atom imaging setup for $^6$Li.} (a) We deterministically prepare a single atom in an optical tweezer and excite it resonantly with two counter-propagating laser beams. With a high-resolution objective (NA=0.55) we collect about 20 fluorescence photons and image them onto an EMCCD camera. (b) Typical raw signal \Note{in analog-to-digital units (ADU)} of a single atom imaged with the EMCCD in electron-multiplication mode (c)  By applying a binarization threshold, we identify pixels where photons impinged. (d) The mean photon signal from a single atom is governed by the random walk of the atom during the photon scattering process and  has a width of $w_\text{sig}=\SI{10.1\pm1.4}{\micron}$. The red arrows indicate the orientation of the probe beams.}
        \label{fig:setup}
\end{figure}



If the requirements on the spatial resolution can be relaxed, the atoms can be imaged based on few fluorescence photons \cite{Buecker2009, Fuhrmanek2010a, Picken2017}. As a consequence, cooling of the atoms is not required, the spin information can become directly accessible, and, under certain conditions, even pinning of the atoms is not necessary. In this paper, we present such a single-atom resolved imaging scheme for fermionic $^6$Li. The experimental scheme is particularly simple and also applicable to other atomic species. In our setup, we achieve single-atom detection fidelities of up to \SI{99.4\pm0.3}{\percent} and a position uncertainty of \SI{4.0\pm0.4}{\micron} based on the detection of about 20 photons per atom. The technique can be applied over a wide range of magnetic fields and allows one to resolve the different hyperfine states in the system. As the atoms are not pinned but propagate freely, the scheme can also be applied after a time-of-flight expansion. Similar to \cite{Buecker2009}, it allows us to access the momentum distribution of a fermionic system with single-atom sensitivity.


\section{Fluorescence imaging of single atoms}
\label{sec:fluor}

Our detection scheme is based on fluorescence imaging of freely propagating fermionic $^6$Li atoms. We drive the D2-transition with resonant light at \SI{671}{\nano\metre} for a few hundred cycles and collect part of the scattered photons with an objective with a large numerical aperture (NA). The collected photons are imaged onto a single-photon sensitive camera, where we can distinguish them from camera noise to identify single atoms. 

With the detection scheme, we image systems of few fermionic $^6$Li atoms that are deterministically prepared in tightly focused optical tweezer potentials \cite{Serwane2011a, Murmann2015a}. The trapping potentials are generated by far-red-detuned Gaussian laser beams at \SI{1064}{\nano\metre} that are focused with the same objective that we use for imaging. To distinguish between different spin states we image the atoms in the range of \SI{300}{} to \SI{1400}{\gauss} in the Paschen-Back regime.



\subsection{Emission and detection of photons}
\label{sec:photem}

In Fig.\@ \ref{fig:setup}(a) we show a schematic of our imaging setup. The high-resolution objective which we use to collect the scattered photons has a numerical aperture of $\textrm{NA}=0.55$ and a focal length of $f=\SI{20.3}{\milli\metre}$. It is chromatically corrected both for \SI{1064}{}-  (trapping) and \SI{671}{}-\si{\nano\metre} (imaging) wavelengths. For the imaging wavelength, the lateral and axial optical resolution is \SI{0.8}{} and \SI{4.4}{\micron}, respectively, and the field of view is approximately \SI{200}{\micron} in diameter. 

We use resonant probe light that is linearly polarized perpendicular to the quantization axis set by the magnetic field orientation ($\hat{\epsilon}\perp\vec{B}$) to drive $\sigma_\pm$ transitions. To mitigate the effect of acceleration due to radiation pressure force, 
 we use two counterpropagating laser beams for the resonant excitation. We avoid detrimental trapping in a standing-wave potential by toggling the two beams alternatingly with pulse lengths of \SI{200}{\nano\second}.
 We aim for a short exposure time and therefore a high photon scattering rate. To simultaneously assure hyperfine-state dependent detection (see Sec.\@ \ref{sec:spinres}), we limit the probe-light intensity to about $I=8.5\,I_\text{sat}$ per beam, where $I_\text{sat}=\SI{2.54}{\milli\watt\per\square\centi\metre}$ is the saturation intensity of $^6$Li. As a result, the natural linewidth of $\Gamma=2\pi\times \SI{5.87}{\mega\hertz}$ \cite{Gehm2003} is power broadened to $\Gamma'=2\pi\times \SI{18.1}{\mega\hertz}$ and the resonant scattering rate is \SI{16.5}{\photons\per\micro\second}.

Good imaging performance at a minimum number of scattered photons requires a high photon collection efficiency. With our high-NA objective, we collect about $\SI{11.4}{\percent}$ of all emitted photons, taking into account the solid angle of the objective (\SI{8.2}{\percent} of the full sphere) and the anisotropic emission pattern due to selectively driven $\sigma_\pm$ transitions. We focus the collected photons on an electron multiplying charge-coupled device (EMCCD, ANDOR iXon DV887, back illuminated) with a quantum efficiency at \SI{671}{\nano\metre} of around $\eta=0.85$ \cite{Becher2016}. We estimate around $\SI{90}{\percent}$ combined transmission through all optical elements in the imaging path, amongst others a band-pass filter centered at \SI{675}{\nano\metre} (Semrock, FF01-675/67-25) to block stray light. As a result, we expect to detect about $\SI{8.7}{\percent}$ of the scattered photons on the camera.

If the atom is not pinned during imaging, it experiences recoils from scattered photons and performs a random walk around the initial position. The diffusive spread of the probability distribution increases with the probe time $\tau$ and decreases the position resolution. For $^6$Li atoms, the recoil energy \si{\er} of \SI{3.5}{\micro\kelvin} leads to a velocity change of about \SI{10}{\centi\metre\per\second} per scattering event and therefore to a much larger diffusion compared to other atomic species. The width of the time-integrated signal increases proportional to $ \tau^{3/2}\sqrt{\Gamma}/(m \lambda)$  \cite{Fuhrmanek2010, Joffe1993}, where $m$ is the mass of the atom and $\lambda$ is the wavelength of the resonant transition. To limit the spread, we choose a relatively short exposure time of $\tau=\SI{20}{\micro\second}$ resulting in about 330 scattered photons. We detect \NoteVV{approximately \num{25} \footnote{The Poissonian photon-number distribution hampers accurate atom-number resolved imaging of multiple coinciding atoms in the same hyperfine state for small average photon numbers. We estimate that number resolution can be implemented above $\sim45$ photons per atom on average \cite{Becher2016})}} photons on average and measure the resulting signal spread governed by the diffusion (see Fig.\@ \ref{fig:setup}(d)) by averaging several hundred images of individual atoms prepared at a fixed initial position. Due to our imaging setup the signal is slightly elliptic and  has a root-mean-square radius of $w_\text{sig}=\SI{10.1\pm1.4}{\micron}$  (see App.\@ \ref{sec:diffusion}). This means that the lateral position resolution in our system is completely dominated by the diffusion process. As a consequence, an axial defocusing of the atoms of up to $\SI{20}{\micron}$ does not deteriorate the single-atom imaging performance.


In general, for a fixed number of scattered photons the width of the integrated photon signal scales as $1/(m \lambda \Gamma)$.  For heavier $^{87}$Rb atoms ($\Gamma=2\pi\times\SI{6.1}{\mega\hertz}$, $\lambda=\SI{780}{\nano\metre}$), the spread for the same number of scattered photons would be reduced by a factor of almost \num{20}. In the case of $^{168}$Er ($\Gamma=2\pi\times\SI{29.7}{\mega\hertz}$, $\lambda=\SI{400}{\nano\metre}$), the reduction is even a factor of about \num{80}.

\subsection{Identification of single atoms}
\label{sec:idatoms}

\begin{figure}[t]
        \centering
        \includegraphics[width=\columnwidth]{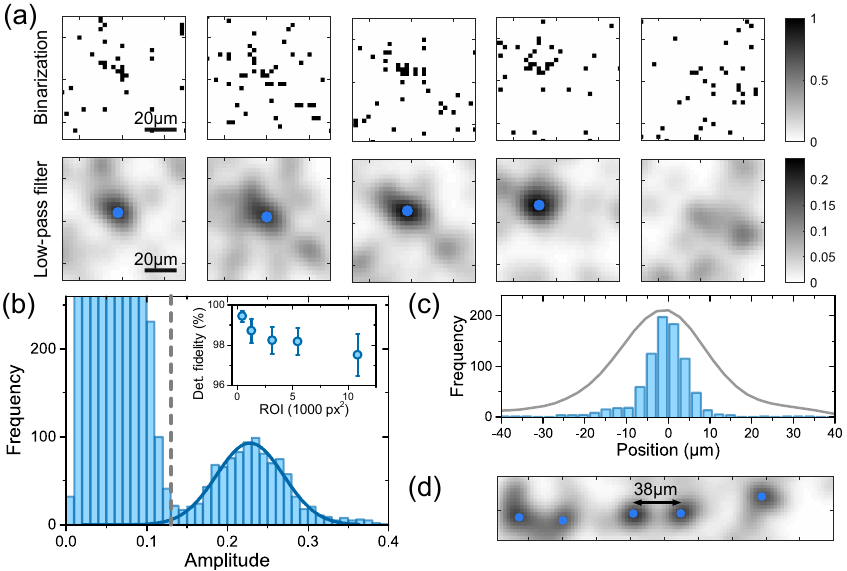}
        \caption{{Performance of the single-atom imaging.} (a) To identify and locate single atoms, we use the EMCCD camera in photon counting mode, apply a low-pass filter to the binarized images and discern single atoms from noise based on the amplitudes of the local maxima in the filtered image. The blue dots correspond to the positions of identified atoms. (b) A histogram of amplitudes shows a bimodal distribution where the peak at lower (higher) amplitude originates from the camera noise (photon clusters). The dashed line depicts the chosen identification threshold for the data taken for a typical image size of $400 \times$ \SI{100}{\micron} corresponding to $\SI{5200}{\square\pixel}$. (b, inset) The detection fidelity depends on the size of the analyzed images and can reach up to \SI{99.4\pm0.3}{\percent}. (c) The position distribution for many realizations of a single atom prepared at a fixed position has an uncertainty of $\sigma_\text{pos}=\SI{4.0\pm 0.4}{\micron}$, which is much smaller than the filter size (gray line). (d) Single image of multiple atoms. The minimal distance at which we can discern individual atoms with above \SI{90}{\percent} probability is around \SI{12}{\pixel} (\SI{32.4}{\micron}). }
        \label{fig:perfim}
\end{figure}


Single-atom detection based on a low photon signal has been demonstrated in \NoteVV{\cite{Miroshnychenko2003, Lester2014, Kwon2017, Picken2017, Hutzler2017}}, where all photons from one atom are focused on a single or few pixels. This ensures that the photon signal exceeds the detector noise. \NoteVV{In our scheme, we aim to detect single atoms with spatial resolution by spreading the signal over several pixels of an EMCCD (see Fig.\@ \ref{fig:setup}(d)) and detecting individual photons in photon counting mode. We use a magnification for which we detect approximately one photon per pixel on average while not diluting the signal too much. This results in compact, prominent clusters of ``bright'' pixels against the background noise (Fig.~\ref{fig:setup}(b) and (c)). The value of the magnification in our system is \num{5.9}, which means that each pixel (\si{\pixel}) corresponds to an area of $2.7 \times \SI{2.7}{\micron}$ in the object plane.}


\NoteVV{The main benefit of the photon counting mode of an EMCCD compared to conventional cameras is that it can detect single photons on each individual pixel (for details see App.~\ref{sec:emccd}).  In a stochastic multiplication process described by an Erlang distribution \cite{Basden2003, Hirsch2013}, the primary photoelectrons contained in a pixel are amplified to several hundred secondary electrons (see legend of Fig.\@ \ref{fig:setup}(b)) such that the amplified signal exceeds the read noise $\sigma_\text{read}$ of the camera. With our camera and the chosen readout parameters, this enhancement is $g/\sigma_\text{read}=64$ where $g$ denotes the average photoelectron gain. We introduce a signal threshold $\sigma_\text{th}$ to discriminate between readout noise and the signal from one or more photons (binarization, see Fig.\@ \ref{fig:setup}(c)).}

In addition to the noise from the readout electronics, EMCCDs suffer from clock-induced charges (CICs) which are randomly distributed and represent the principal noise source of the single-photon detection process. We can suppress a fraction of the CICs \cite{Lantz2008} by choosing a high threshold of $\sigma_\text{th}=8 \, \sigma_\text{read}$ and thereby reduce the contribution of CICs to about \SI{1.7}{\percent} while identifying \SI{88.5}{\percent} of the photoelectrons (see App.~\ref{sec:emccd}).
 
Figure \ref{fig:perfim}(a), top row, shows typical images of the photon signals after binarization. The clusters in the center arise from the fluorescence of single atoms and the randomly distributed noise is caused by the CICs. 
To identify the signal from a single atom, we make use of the different spatial frequency spectra of the atom signal and the noise. 
\Note{After applying a Gaussian low-pass filter (3-\si{\pixel} or 8.1-\si{\micron} width), photon clusters lead to local maxima with much higher amplitude than the local maxima caused by the CICs (Fig.\@ \ref{fig:perfim}\,a, bottom row).}

\Note{We record the amplitudes of these local maxima in a histogram, which contains the data from several hundred images with at most a single atom per image (Fig.\@ \ref{fig:perfim}(b)). In the resulting bimodal distribution, we identify the peaks at higher and lower amplitude as being caused by clusters of fluorescence photons and CICs, respectively. \NoteV{Based on this, we introduce an identification threshold above which we identify a peak in the image with the presence of a single atom (blue dots in Fig.~\ref{fig:perfim}(a)).} \NoteVV{We choose this threshold value as the amplitude for which the numbers of false positive events (detection of an atom when none are present) and false negative events (an atom is present, but not detected) in an image are equal.} \footnote{We obtain the probability of false negatives at a certain identification threshold from a histogram over images which contain no atoms. The probability of false positives is extracted from a Gaussian fit as indicated in Fig. 2(b).}. The detection fidelity is then defined as the probability of true positive events.}


\Note{With decreasing overlap between the two peaks of the bimodal distribution, the probability of false identifications of either kind decreases. As the total amount of CICs scales with the image size, this overlap and consequently the detection fidelity depends on the image size, as shown in Fig.\@ \ref{fig:perfim}(b, inset).  For small image sizes of $21\times$\SI{21}{\pixel}, we reach a single-atom detection fidelity of up to \SI{99.4\pm0.3}{\percent}.}

By extracting the position of the local maximum in a filtered image, as shown in Fig.\@ \ref{fig:perfim}(a), we also determine the position of a single atom. However, the peak position of the signal does not directly correspond to the initial position of the atom due to the random walk. We experimentally determine the position uncertainty by repeatedly imaging individual atoms which have been prepared in a single tweezer at a fixed position. This ensures that the atom's initial position is known prior to imaging. From the distribution of the determined positions for several hundred repetitions (see Fig.\@ \ref{fig:perfim}(c)), we obtain a position uncertainty of $\sigma_\text{pos}=\SI{4.0\pm 0.4}{\micron}$\NoteVV{, corresponding to one standard deviation}.

We can also analyze images of multiple atoms with our identification scheme (Fig.\@ \ref{fig:perfim}(d)). The minimal distance at which we can discern two nearby atoms from each other depends on the magnitude of the diffusion and the chosen size of the lowpass filter. We determine this distance experimentally by preparing two atoms with the same hyperfine state and letting them expand (see Sec.\,\ref{sec:kspace}). From the distribution of inter-atomic distances we extract a minimal distance of about \SI{12}{\pixel} (\SI{32.4}{\micron}) above which we can identify two nearby atoms \NoteV{in the same hyperfine state} with a probability of over \SI{90}{\percent}. 



The main factor limiting the fidelity of our detection scheme is the camera noise from spurious charges. Modern EMCCD cameras show a significantly reduced occurrence of CICs\NoteVV{, typically on the order of \SI{0.3}{\percent} at the same photon detection efficiency. Therefore, one could achieve the same atom detection fidelities from fewer scattered photons.} This makes it possible to reduce the exposure time and thereby improve the \NoteV{position resolution of the imaging.} \NoteV{The position resolution can be further improved for \textit{in-situ} measurements by pinning the atoms in optical tweezers during imaging (see App.\,\@\ref{sec:diffusion}), similar to other imaging schemes with optical tweezers \cite{Lester2014, Kim2016, Kwon2017, Picken2017, Hutzler2017}.}



\section{Spin resolution}
\label{sec:spinres}

\begin{figure}
        \centering
                \includegraphics[width=\columnwidth]{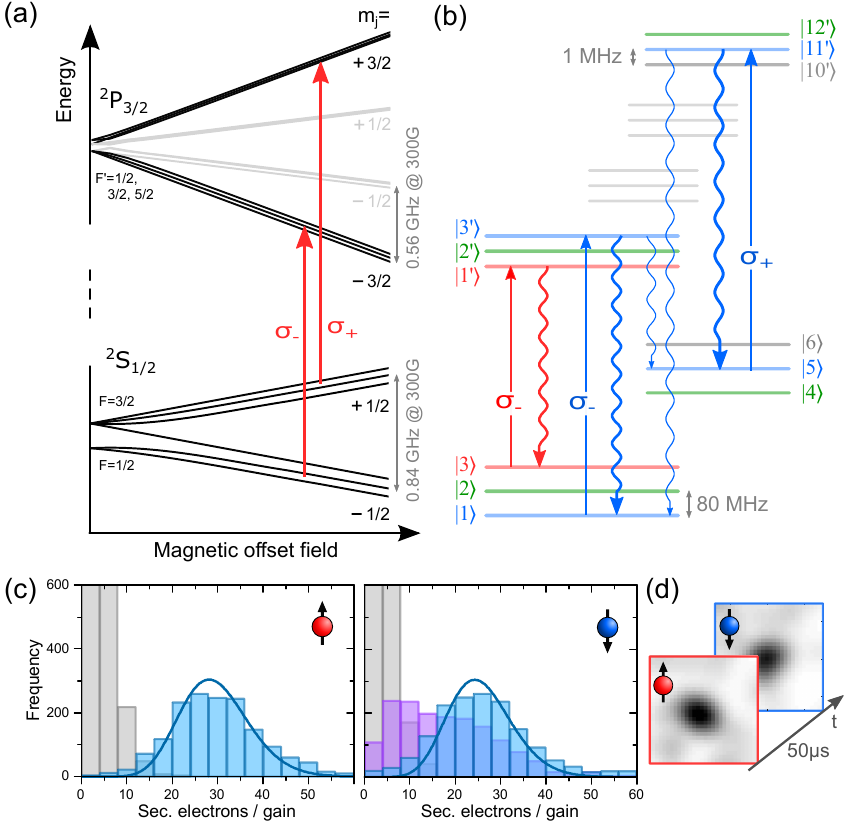}
        \caption{{Spin-resolved imaging of $^6$Li. (a) At a non-zero magnetic field, we drive the $\sigma_-$ transition from the $m_j=-1/2$ to $-3/2$ manifold and the $\sigma_+$ transition from the $m_j=+1/2$ to $+3/2$ manifold which all contain $m_I=\pm 1, 0$ sublevels. (b) As the optical frequencies differ by \SI{80}{\mega\hertz}, we can excite the three lowest hyperfine states $\ket1$, $\ket2$, and $\ket3$ selectively to the three lowest hyperfine states of the excited manifold $\ket{3'}$, $\ket{2'}$, and $\ket{1'}$. With more than \SI{99}{\percent} probability, the atoms then decay back to their initial state. To compensate for the small probability that atoms in $\ket{2'}$ and $\ket{3'}$ can decay to $\ket{4}$ and $\ket{5}$, respectively, we additionally drive the $\sigma_+$-transition. (c) Distribution of the integrated signal from a single atom in state $\ket3$ probed with a closed optical transition (left panel) and in state $\ket1$ (right panel) at \SI{568}{\gauss} without (purple) and with (blue) the additional $\sigma_+$ probe beam. The gray bars show the signal with no atom present, solely caused by CICs. We model the distribution (solid line) with a mean photon number of 24.7 and 21.0 for state $\ket 3$ and $\ket 1$, respectively. (d) We achieve spin-resolved imaging of two atoms in different hyperfine states by subsequently addressing their resonant transition and using the kinetics mode of the camera.}}
        \label{fig:second_graph}
\end{figure}


\Note{In two-component many-body systems, \NoteV{atoms in different spin states may be correlated.} Single-atom imaging with spin resolution can shed light on the presence of such correlations. Different hyperfine states in a system can be discerned if they can be addressed in a selective manner\NoteV{, for example by optical transitions with different resonance frequencies. To obtain spin resolution, mixing of the initially populated hyperfine states during imaging has to be avoided. Also, fluorescence imaging in general requires that the respective optical transitions are closed (cycling transitions). Otherwise, scattering several hundred photons would lead to a significant probability that the atom decays into an off-resonant state. This would reduce the average number of scattered photons from a single atom \cite{Klinkhamer2018} and, consequently, the fidelity of the single-atom detection. Both the selective addressability and the existence of cycling transitions are determined by the electronic structure of the atom.}}


\NoteV{In $^6$Li, applying a magnetic offset field during imaging allows us to selectively address different hyperfine state, while also minimizing detrimental decay into dark states.} $^6$Li has one valence electron and a nuclear spin of $I=1$. Consequently, the levels split according to the $z$ component of the total angular momentum $m_j$ and the nuclear spin $m_I$ in the magnetic field, as shown in Fig.\@ \ref{fig:second_graph}(a). As the hyperfine coupling constant is relatively small \cite{Gehm2003}, $I$ and $J$ decouple already above \SI{30}{\gauss} for the $^2S_{1/2}$ state and above \SI{1}{\gauss} for the $^2P_{3/2}$ state.  We work with two of the three lowest hyperfine states of $^6$Li, denoted by $\ket 1$, $\ket2$ and $\ket3$, ordered ascending in energy.  At large magnetic fields, the frequencies of their optical transitions differ by about \SI{80}{\mega\hertz}, which is a factor of 12 larger than natural linewidth $\Gamma$. This allows us to address the hyperfine states individually.

To probe the atom on a cycling transition, we use resonant light with linear polarization perpendicular to the quantization axis. \NoteV{ In that way, we resonantly drive a $\sigma_{-}$ transition ($\Delta m_j=-1$) from the $m_j=-1/2$ to the $m_j'=-3/2$ manifold. From there, only a decay back to $m_j=-1/2$ is possible, as the nuclear spin projection $m_I$ is conserved under photonic excitations. $\sigma_{+}$ transitions from the $m_j=-1/2$ to the $m_j'=+1/2$ manifold are strongly suppressed due to the detuning by more than \SI{1.1}{\GHz} above \SI{300}{\gauss} and can be neglected.} 

\Note{Due to angular momentum coupling of $I$ and $J$, each state $\ket{m_j, m_I}$ is approximately an eigenstate of the total angular momentum operator of the electron $\hat J$  only in the limit of high magnetic field.} \NoteVV{Residual couplings lead to small admixtures of different spin components $m_j$ and result in losses from a cycling transition into dark states.}

\NoteVV{In the hyperfine state $\ket 3$, the atom can be exactly described by a single total angular momentum state  $\ket{m_j=-1/2, m_I=-1}$ without additional admixtures. It is a stretched state and has a perfectly closed optical transition to the excited state $\ket 1' = \ket{-3/2, -1}$ (see Fig.\@ \ref{fig:second_graph}(b)). The probability that the atom becomes dark during imaging is therefore zero, and we detect approximately \num{25} photons on average within \SI{20}{\micro\second},} as shown in Fig.\@ \ref{fig:second_graph}(c, left panel).

The hyperfine states $\ket1$ and $\ket2$\NoteV{, however,} contain a small admixture of $\ket{m_j=+1/2}$ \cite{Gehm2003}. 
After an excitation, the probability that an atom decays into states $\ket5$ and $\ket4$, respectively, is on the order of a few permille above \SI{500}{\gauss} \cite{Becher2016}. We circumvent the loss into a dark state by adding a second optical frequency to the probe beam, which resonantly drives the $\sigma_{+}$ transition from $m_j=+1/2$ to $m_j'=+3/2$ (see Fig.\@ \ref{fig:second_graph}(b)). Atoms that have decayed to the states $\ket 4$ and $\ket 5$ are resonant to the second frequency \Note{of this bichromatic probe beam} and continue to scatter photons. 
\Note{\footnote{With the bichromatic probing scheme, discerning the hyperfine states $\ket1$ and $\ket5$ is  not possible.}}
The increased number of detected photons is clearly visible from the signal distribution in Fig.\@ \ref{fig:second_graph}(c, right panel) which corresponds to in average \num{21.0} scattered photons. We reach a detection fidelity of $\SI{97.8\pm0.7}{\percent}$ for the states $\ket1$ and $\ket2$ \Note{using the same analysis as explained in Sec.\@ \ref{sec:idatoms}}. The residual loss probability from other effects like polarization misalignment or off-resonant transitions is below $2\times10^{-4}$ per scattering event.


To image atoms in the different hyperfine states, we successively probe the individual optical transitions. We resonantly drive one hyperfine state for \SI{20}{\micro\second} and image the scattered photons on the EMCCD camera. \NoteVV{We then shift the laser frequency to the resonance frequency of the other hyperfine state as fast as possible and simultaneously use the fast kinetics mode of the camera to shift the collected signal out of the exposed sensor area (\SI{0.5}{\us} per line).} \Note{For our typical regions of interest, this takes about \SI{30}{\micro\second}}. After that, we image the second hyperfine state for \SI{20}{\micro\second}. In that way, we can take separate images of the different hyperfine states, as illustrated in Fig.\@ \ref{fig:second_graph}(d).
\NoteVV{In principle, we can use this method to successively image an arbitrary number of spin states.} \Note{For example, in a three-component system, we can image all three lowest hyperfine states $\ket 1, \ket 2$, and $\ket 3$.} 

\Note{For the successive probing, we must also ensure that the typical dynamics of the system during imaging are sufficiently slow. Only then are we able to measure the same observable for both spin states. The timescale of the system strongly depends on the specific implementation; for expanding atoms, for example, it may be given by the Fermi velocity.} Furthermore, off-resonant scattering of photons during the first imaging pulse can broaden the distribution in the second image \Note{and thereby decrease the resolution}. In our measurements, we observe that  in the direction of the probe beams the position uncertainty in the second image is \SI{10}{\percent} larger than in the first image which indicates off-resonant scattering of one photon on average.




\section{Single-atom imaging in momentum space}
\label{sec:kspace}

\begin{figure}
        \centering
                \includegraphics[width=\columnwidth]{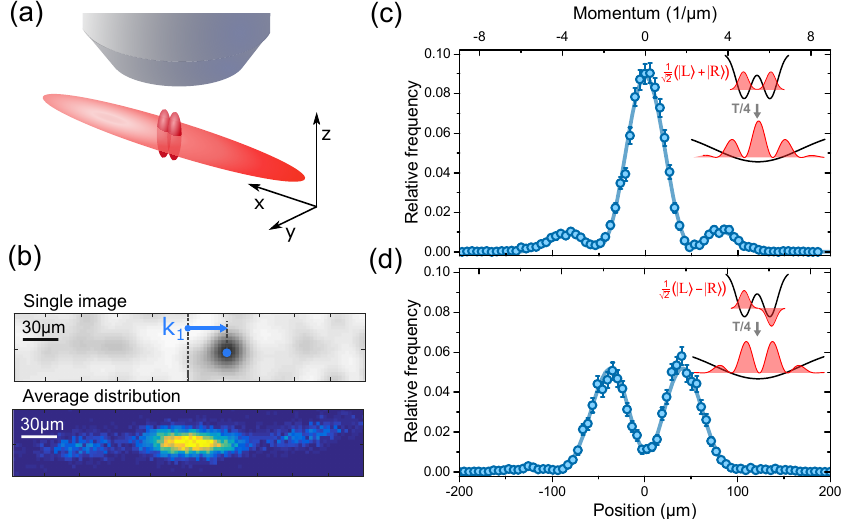}
        \caption{{Momentum imaging of single atoms. (a) The double-well potential formed by two overlapping optical tweezers is centered at the optical dipole trap elongated along the {\it x} axis. The latter provides a harmonic potential for the expansion of the quantum state. (b) After an expansion for a quarter trap period along the {\it x} axis, $t=T_{x}/4$, the initial momentum wave function is mapped to real space and the average distribution of the atom corresponds to the momentum distribution of the state, shown here for a single atom prepared in the ground state of the double-well potential.  (c) We detect its position distribution after the expansion and observe a probability distribution as expected from a double-slit experiment with constructive interference at the center. (d) Preparing the anti-symmetric superposition of $\ket L$ and $\ket R$ results in destructive interference at the center.
 }}
        \label{fig:fourth_graph}
\end{figure}

Our imaging scheme can also be applied to freely propagating atoms, e.g., during time of flight. For sufficiently long expansion times, this allows us to access the momentum distribution of the initial state on a single-atom level. To detect the atoms with high fidelity, they have to be in the focal plane of the objective during imaging. Therefore, the expansion of the quantum state wave function is performed in a \Note{confining optical potential} that prevents the atoms from leaving the focal plane of the objective.
If this potential is harmonic along the focal plane, an expansion for a quarter trap period $T/4$ exactly maps the momentum distribution of the initial state to position space \cite{Amerongen2008, Jacqmin2012, Murthy2014}.


\Note{In our setup, we make use of this momentum-space imaging along one direction. Our confining optical potential is formed by an elongated crossed-beam optical dipole trap with an aspect ratio $\eta \approx 8$  where the long axis lies in the focal plane (see Fig.\@ \ref{fig:fourth_graph}, a). By expanding the system for $T/4$ of the trap period along the long axis, we image the momentum distribution along the elongated axis, while integrating along the other two axes. } We choose the depth of this trap to be as weak as possible while still confining the atoms in the harmonic region of the trap during the expansion. The resulting trap frequency in the axial direction on the order of $\omega_\text{wg}=2 \pi \times\SI{75}{\hertz}$ leads to an atom distribution over about \SI{100}{\micron}, which is significantly larger than the single-atom resolution (see Fig.\@ \ref{fig:fourth_graph}(b)). As a consequence, we can measure the momentum of a single atom with a \Note{momentum uncertainty of $\sigma_\text{mom}=\SI{0.18}{\per\micron}$}. 

We demonstrate the momentum resolution by imaging a single atom that has been deterministically prepared in an eigenstate of the double-well potential created with two optical tweezers. The axis of the double-well potential coincides with the long axis of the confining potential (Fig.\@ \ref{fig:fourth_graph}(a)). As shown in earlier works \cite{Murmann2015a}, we deterministically prepare a single atom in the ground state of the symmetric double-well potential, which can be described as a symmetric superposition of the two Wannier functions $\ket L$ and $\ket R$. Then we perform the expansion of the quantum state for $T/4$ and image the atom. Figure \ref{fig:fourth_graph}(c) shows the distribution of the detected positions of the single atom after the expansion. In analogy to a double-slit experiment, it shows a fringe pattern with constructive interference at the center and a spacing that is in agreement with the double-well separation. Preparing the first excited state of the double-well potential, which is the antisymmetric superposition of $\ket L$ and $\ket R$, we observe a similar pattern with destructive interference at the center (see Fig.\@ \ref{fig:fourth_graph}(d)).

\section{Summary and Outlook}
We have presented an imaging method to characterize few-particle quantum states of $^6$Li on a single-particle level with spin and position resolution. We have demonstrated single-atom imaging by collecting about 20 fluorescence photons per atom on a single-photon sensitive camera. With this method we achieve a detection fidelity of up to \SI{99.4\pm0.3}{\percent} and a position determination with an uncertainty of \SI{4.0\pm0.4}{\micron}. Our imaging technique has been applied to freely propagating atoms also after a coherent expansion in time-of-flight. We demonstrate the detection of momentum distribution on the single-atom level by an expansion in an \Note{ elongated confining trap}.

In the future, we will apply the free-space imaging also to systems containing more atoms and after more complex manipulations of the system. For example, a sequence of coherent expansions in different harmonic potentials could be used to magnify the quantum state in position space. In that way, we could access the {\it{in situ}} distribution of atoms in multiple optical tweezers with single-atom sensitivity. This opens a path to studying correlation functions of complex quantum systems in two conjugate bases, real space and momentum space.

\section*{Acknowledgment}
We thank R.\@ Rosa-Medina, J.\@ Niedermeyer, P.\@ L.\@ Bommer, L.\@ J.\@ Thorm\"alen, P.\@ Lysakovski, and T.\@ Steinle for contributions at an early stage of the experiment;  P.\@ A.\@ Murthy for careful reading of the paper and M.\@ Storath for fruitful discussions on the image analysis. We thank the group of T.\@ Pfau for lending equipment. A.\@ B.\@ acknowledges funding from the International Max Planck Research School for Quantum Dynamics in Physics, Chemistry and Biology.  P.\@ M.\@ P.\@ acknowledges funding from European Union's Horizon 2020 program under the Marie Sk\l odowska-Curie Grant No.\,706487. This work has been supported by the European Research Council consolidator grant 725636 and is part of the Deutsche Forschungsgemeinschaft Collaborative Research Centre ``SFB 1225 (ISOQUANT)''.

A.\@ B. and V.\@ M.\@ K. contributed equally to this work.

\appendix

\section{Readout of the EMCCD camera}
\label{sec:emccd}

\begin{figure}
        \centering
                \includegraphics[width=\columnwidth]{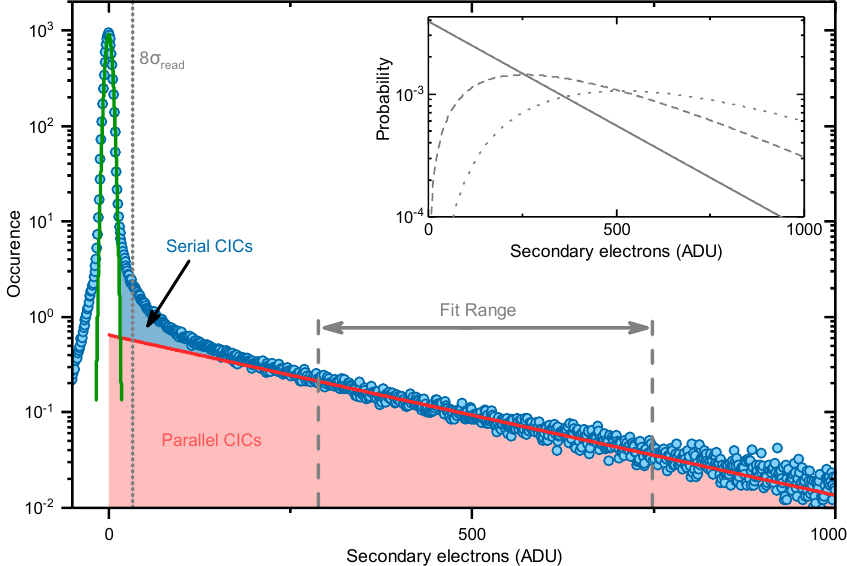}
        \caption{\Note{Logarithmic histogram of secondary electrons on a dark image. With the EMCCD camera stochastic amplification of few photoelectrons is possible to overcome the read noise. The distribution of secondary \NoteVV{electrons} is shown (inset) for one (solid line), two (dashed line), and three (dotted line) primary electrons. We extract the histogram of secondary electrons on a bias-corrected image with a closed camera shutter (blue dots). From this, we extract the read noise $\sigma_\text{read}$ by fitting a Gaussian function (green solid line) to the peak around zero. The exponential tail at higher values of secondary electrons is caused by clock-induced charges (CICs) from the shifting processes in the EMCCD, where parallel CICs (red region) and serial CICs (blue region) can be distinguished. We fit the pCICs by a simple exponential function (red solid line) and extract the average gain $g$. Due to a significant fraction of sCICs, we choose a binarization threshold of $8\sigma_\text{read}$ (gray dotted line) to discern a photon signal from read noise.}}
        \label{fig:emccd}
\end{figure}

\Note{To detect the fluorescence photons with high fidelity, we use an EMCCD (ANDOR iXon DV887). During the read-out of an image, the photoelectrons created by impinging photons are first shifted row by row to a shift register. Then, they are transferred into the gain register where the charges are amplified in several hundred stages. The amplification process of $n$ primary photoelectrons to $x$ secondary electrons is stochastic and \NoteVV{its probability} can be described by the Erlang distribution (see Fig.\@ \ref{fig:emccd}, inset) \cite{Lantz2008, Harpsoe2012}
\begin{equation}\label{eq:erlang}
P_n (x)=\frac{x^{n-1} \mathrm{e}^{-x/g}}{g^n (n-1)!} \theta(x)\,\text{,}
\end{equation}
where $g$ denotes the average gain per photoelectron \NoteVV{and $\theta(x)$ is the Heaviside step function}. For a single primary photoelectron ($n=1$), the distribution of secondary electrons reduces to an exponential function. \NoteVV{This single photoelectron can only be detected if the number of secondary electrons that it produces significantly exceeds the read noise $\sigma_\text{read}$, e.g., by using a large gain $g$.} \NoteV{In this detection mode, the stochastic nature of the amplification does not allow for precise photon counting on each pixel.}}

\Note{We can characterize the camera by analyzing images without photons. A histogram of the signal on all pixels after a bias correction is shown in Fig.\@ \ref{fig:emccd} where the peak centered at zero secondary electrons (given in analog-to-digital units) corresponds to the normally distributed read noise. The tail at large signals is due to CICs, which are the primary noise source of the EMCCD camera. The so-called parallel CICs (pCICs) are charges created on the sensor by the parallel shifting process, and therefore indistinguishable from photoelectrons. In the multiplication register, they are amplified with a fixed gain following Eq.\@ (\ref{eq:erlang}). The serial CICs (sCICs) are created at random stages within the serial multiplication register and have a variable gain that is stochastically distributed \cite{Lantz2008}. In Fig.\@ \ref{fig:emccd}, they are visible as a shoulder (blue shaded area) between the read-noise peak and the exponential distribution from the pCICs. To quantify the performance of the EMCCD in photon counting mode, we fit the Gaussian read noise and the exponential distribution of pCICs to obtain $g/\sigma_\text{read}=64$.}

\Note{For faint signals where at most one photon impinges per pixel, the amplitude of a pixel conveys no additional information. In photon counting mode, we divide the pixels into bright and dark pixels above or below a threshold $\sigma_\text{th}$, respectively (binarization). In that way, we can suppress the read noise of the camera in the limit of large $g/\sigma_\text{read}$ at the expense of only a small reduction of the single-photon detection fidelity to $\exp(-\sigma_\text{th}/g)$. The residual noise on a dark image is due to CICs. As the sCICs are significant only at low numbers of secondary electrons, we can suppress a large part of the noise by choosing a binarization threshold at $8 \,\sigma_\text{read}$. We end up with \SI{1.7}{\percent} of spurious bright pixels due to CICs and a photon detection efficiency of \SI{88.5}{\percent}.}



\begin{figure}
        \centering
                \includegraphics[width=\columnwidth]{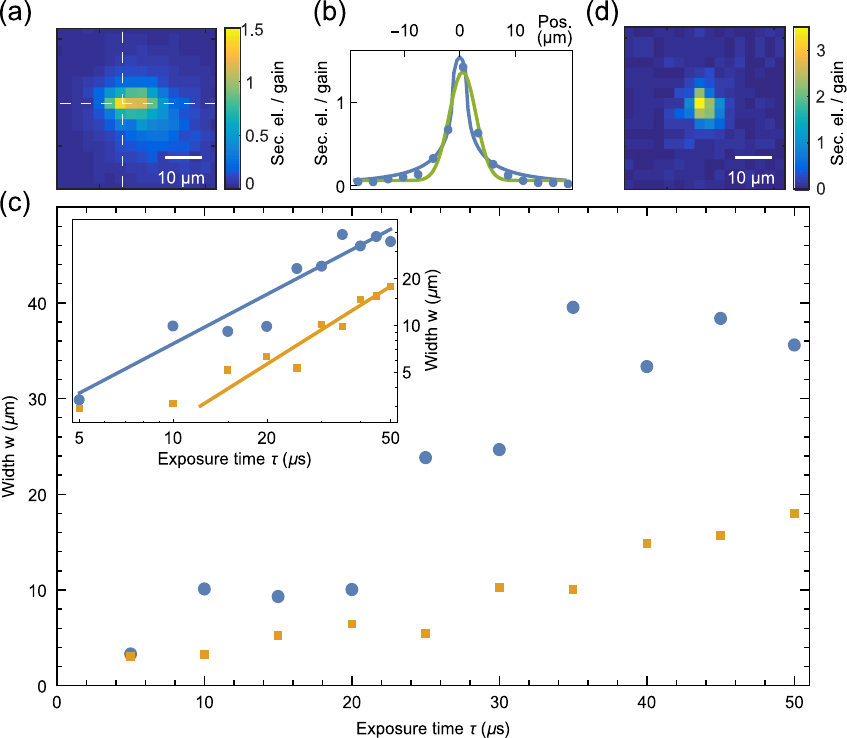}
        \caption{(a) Average photon distribution after an exposure time of \SI{20}{\us}. $x$ and $y$ profiles are extracted along the dashed lines. (b) $y$ profile of the photon distribution (blue points), fitted with $\tilde{\rho}(0,y_i,\tau)$ (blue curve) and with a Gaussian distribution (green curve). (c) Width $w = \sqrt{s}$ as extracted from the fitted model vs.\@ exposure time, for free expansion (blue points) and with a pinning potential with a depth of \SI{290}{\er} (yellow squares). Inset: Double-logarithmic plot of the width, showing the scaling behavior of the expansion (lines are guides to the eye). (d) Average photon distribution after an exposure time of \SI{20}{\us} with a pinning potential with a depth of \SI{290}{\er}.}
        \label{fig:third_graph}
\end{figure}

\section{Diffusion during imaging}
\label{sec:diffusion}

When scattering photons, an atom performs a random walk in momentum and position space, due to the randomness of the spontaneous emission. Its position distribution is described by a normal distribution with a width growing as $\sigma^2(t)=\frac{R \alpha}{3} v_\text{rec}^2 t^3 \equiv 3 \delta^2 t^3$ \cite{Joffe1993}, with the scattering rate $R$, the recoil velocity $v_\text{rec}$, and the dipole pattern correction factor $\alpha$. If we image an atom with resonant light for an exposure time $\tau$, the atom will diffuse over the course of the imaging process. Hence, the detected photon distribution arises from the time-averaged position distribution of the atom,
\begin{equation}
\rho(x,y,\tau) = \frac{1}{6 \pi \delta^2 \tau^3} \, E_{1/3}\!\!\left(\frac{x^2 + y^2}{2 \delta^2 \tau^3}\right)\text{,}
\label{eq:photondist}
\end{equation}
where $E_n (z) = \int_1^\infty \frac{e^{-z u}}{u^n} \text{d}u$ is the generalized exponential integral.

We compare this model to the experimentally obtained average image of single atoms (Fig.\@ \ref{fig:third_graph}(a)). Knowing that our model only includes the effect of spontaneous emission on the atom distribution, we introduce a variable width factor $b$ (by substituting $x \rightarrow b x$) to effectively absorb the effects of other processes. Also, we integrate the distribution over the area of each pixel $i$, resulting in $\tilde{\rho}(x_i,y_i,\tau) = \int_{x_i-\si{\pixel}/2}^{x_i+\si{\pixel}/2}\int_{y_i-\si{\pixel}/2}^{y_i+\si{\pixel}/2} \rho(b x,b y,\tau) \, \text{d}x \text{d}y$, where $x_i$ and $y_i$ stand for the coordinate of the pixel. We then use this model to fit different profiles of the average image (Fig.\@ \ref{fig:third_graph}(b)). Compared to a fitted normal distribution (green), the model (blue) performs systematically better in the wings of the distribution.

We can now use the model to characterize our imaging for different exposure times. 
As a figure of merit, we use the width $w(\tau) = \sqrt{s(\tau)}$ obtained from the second moment of the fitted curves, $s(\tau) = \sum_i x_i^2 \tilde{\rho}(x_i,\tau)$. For exposure times between \SI{5}{} and \SI{50}{\us}, we observe that the diffusion widens the observed distribution significantly already for short exposure times (blue points Fig.\@ \ref{fig:third_graph}(c)). Hence, it is possible to optimize the position resolution by minimizing the exposure time, while retaining a sufficient signal-to-noise ratio for atom identification. For the free-space measurements described in this paper, we limited the exposure time to \SI{20}{\us}, which resulted in a signal of approximately \num{20} photons per atom on average and a resolution of \SI{4.0(4)}{\micron}.

The position resolution can also be improved by pinning the atoms in a deep potential during imaging, which inhibits their random walk. For a single atom in a single microtrap, we have tested this method up to a depth of \SI{290}{\er} (Fig.\@ \ref{fig:third_graph}(d)). \NoteVV{At the maximum depth, we see that the distribution starts broadening only after approximately 10 to \SI{20}{\us} of exposure time (yellow points in Fig.\@ \ref{fig:third_graph}(c)), which corresponds to \numrange[range-phrase = --]{160}{330} scattered photons. This suggests that the atom has initially been confined in the tweezer during imaging.} For longer exposure times, the width starts to increase at a similar rate as the unconfined atoms from the previous paragraph (blue and yellow solid lines in the inset of Fig.\@ \ref{fig:third_graph}(c)), \NoteVV{indicating that the atom has been heated out of the tweezer. To pin atoms in an array of microtraps for an exposure time of \SI{20}{\us} (\num{330} scattered photons), we estimate that we need at least \SI{40}{\mW} of optical power per tweezer (corresponding to a depth of \SI{330}{\er}). As an added benefit, an increased depth of the tweezers will also improve the position resolution.}


\bibliographystyle{apsrev4-1}

%

\end{document}